\documentclass[journal, twoside]{IEEEtran}
\usepackage{textcomp}
\usepackage{amsmath,amssymb,amsfonts}
\usepackage{algorithmic}
\usepackage{graphicx}
\graphicspath{{./Figures/}}
\DeclareGraphicsExtensions{.pdf,.jpeg,.png}
\usepackage{array}
\usepackage{wrapfig}
\usepackage{subcaption}
\usepackage{stfloats}
\usepackage{verbatim}
\usepackage{rotating}
\usepackage[style=ieee, backend=bibtex]{biblatex}
\usepackage{ifthen}

\makeatletter
\newcounter{IEEE@bibentries}
\renewcommand\IEEEtriggeratref[1]{%
	\renewbibmacro{finentry}{%
		\stepcounter{IEEE@bibentries}%
		\ifthenelse{\equal{\value{IEEE@bibentries}}{#1}}
		{\finentry\@IEEEtriggercmd}
		{\finentry}%
	}%
}
\makeatother

\newcolumntype{L}{>{\raggedright\arraybackslash}p}
\newcolumntype{C}{>{\centering\arraybackslash}m}

\begin{document}

\title{A Passive Re-Directing Van Atta Type Reflector}
\author{Paris Ang and George V. Eleftheriades
\thanks{This work has been supported by the Natural Sciences	and Engineering Research Council of Canada (NSERC) and the Ontario Centers of Excellence (OCE).}
\thanks{The authors are with the Edward S. Rogers Sr. Department of Electrical and Computer Engineering, University of Toronto, Toronto, ON M5S 3G4, Canada	(e-mail: paris.ang@mail.utoronto.ca; gelefth@waves.utoronto.ca).}
}

\maketitle

\begin{abstract}
\boldmath
The Van Atta retro-reflector can be envisioned as a self-phasing antenna array; a phase gradient, naturally induced by an external incident wave, is inverted so the device re-radiates back towards the incident direction. This Letter demonstrates how to re-direct the re-radiated beam through passive alteration of this phase gradient. For this purpose, cross-propagating isolation is required between the incident and re-radiated signal paths. To this end, polarization duplexing can be used to achieve this isolation with a passive and reciprocal structure. To provide further demonstration, a 4-element re-directive array is designed and fabricated to offset its re-radiated beam by $-15^{\circ}$ from incidence. Performance is then verified using fullwave electromagnetic simulation and experimental radiation pattern measurements. It is found that the $-15^{\circ}$ angular offset is effectively retained over a wide range of incidence angles.
\end{abstract}


\section{Introduction}
\label{sec:introduction}
A modification is proposed herein to passively alter the re-radiation angle of a Van Atta retro-directive array \cite{Atta1959}. Here, the re-radiated beam is steered to a desired angle by imparting an additional phase gradient into the received signal. While the idea of changing a Van Atta array's output angle is not new, previous proposals have focused on active techniques like phase-conjugating mixing \cite{Miyamoto2002} or require non-reciprocal elements \cite{Davies1963}. In contrast, the proposed technique can be performed passively and without the use of non-reciprocal elements, external power, or control signals.

Passive beam re-direction has also become a popular focus in recent metasurface research with applications in communications and radar cross-section (RCS) reduction \cite{Modi2017}, \cite{Li2014}. These can be fabricated to achieve highly efficient wavefront retro-reflection \cite{Wong17}, \cite{Estakhri2017} and re-direction \cite{Diaz-Rubioe2017}, \cite{Epstein2016}. However, current re-direction metasurface designs only operate effectively at a single set of incidence and re-radiation angles. In contrast, the proposed method consistently applies the same angular offset over a wide range of incident angles at a fixed frequency.

As a demonstration, a 10 GHz patch antenna based reflector was designed to re-radiate a beam offset $-15^{\circ}$ with respect to incidence. The reflector's performance was then assessed through simulation and experimental measurements.

\section{Re-Directive Array Theory}
\label{sec:theory}
\begin{figure}[!tb]
	\centering
	\includegraphics[width=3.25in]{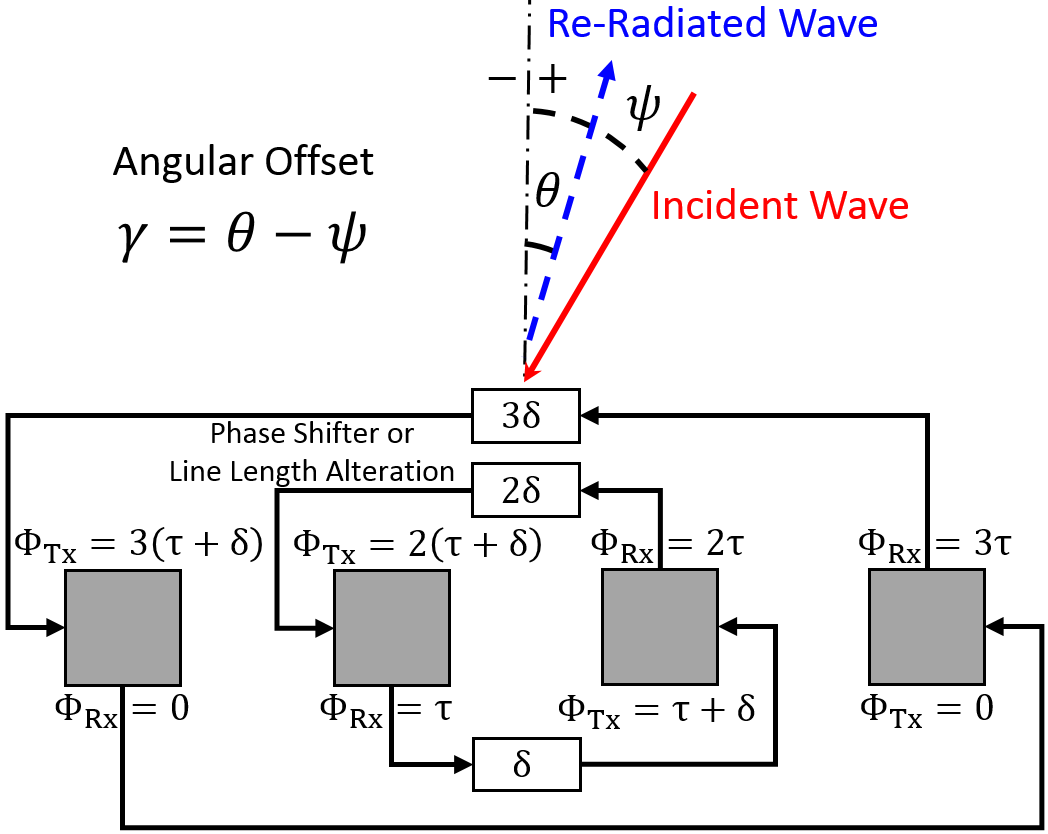}
	\caption{Polarized-duplexed re-directing reflector}
	\label{fig:VRPol} 
\end{figure}

A uniformly spaced N-element phased array is capable of controlling its direction of radiation by maintaining a phase gradient $\alpha=-k_{0}d\sin{\theta}$ between each element \cite{Balanis2005}. Here, $k_{0}=2\pi/\lambda_{0}$ is the free space wavenumber, $d$ is the inter-element spacing, and $\theta$ is the desired radiation angle referenced to the array's normal axis. The input phase for each element ($\phi_{Tx}$) is thus an increasing multiple of this gradient.

A Van Atta retro-reflector can be considered an array that attains its power and phase gradient from a received incident wave and interconnecting lines. Due to differences in propagation distances, a wave with an off-boresight incident angle of $\psi$ is captured with a phase gradient of $\tau=-k_{0}d\sin{\psi}$. Usually, both elements in each pair receive and transmit simultaneously; capturing and feeding electromagnetic energy to its partner via interconnecting lines to be re-radiated. If the lines are of equal lengths, or differ by multiples of the guided wavelength, the phase gradient is inverted when the signals reach their partner antennas and the array radiates towards the source. As the phase gradient is induced externally, only one connecting line is required per pair \cite{Larsen1964}.

Consequently, the direction of re-radiation can be altered by changing this naturally induced phase gradient. Fig. \ref{fig:VRPol} details such a concept where an additional phase gradient $\delta$ can be passively imparted with phase shifters or by adding or removing segments of interconnecting lines. The total transmitted phase gradient then becomes:
\begin{equation}
	\alpha=\tau+\delta=-k_{0}d\sin{\psi}+\delta
\end{equation}
resulting in a re-radiated beam angle of:
\begin{equation}
	\theta=\frac{-\arcsin(\alpha)}{k_{0}d}
\end{equation}
Thus, the additional phase gradient for a re-radiation angle of $\theta$ when the array is illuminated at incident angle $\psi$ is:

\begin{equation}
	\delta=k_{0}d\left(\sin{\psi}-\sin{\theta}\right)
	\label{eqn:SteeringAngle}
\end{equation}

To achieve re-direction, the added phase must be an increasing multiple of $\delta$ at each element. This requires \textit{isolated connecting networks so a separate phase offset can be imparted to each counter-propagating wave}. While the use of non-reciprocal circuits like circulators has been suggested \cite{Davies1963}, a simpler solution is proposed in Fig. \ref{fig:VRPol}. In this design, pairs of dual polarized antennas are connected with two lines and polarization duplexing is exploited to achieve cross-propagating isolation. A multiplicative phase shift can then be added to each line to augment the phase gradient. The advantage of this design is that no non-reciprocal or active components are required. However, each element pair now requires two connecting lines, increasing routing complexity.

\section{Design and Simulation}
\label{sec:simulation}
\begin{figure}[!tb]
	\centering
	\includegraphics[width=3.25in]{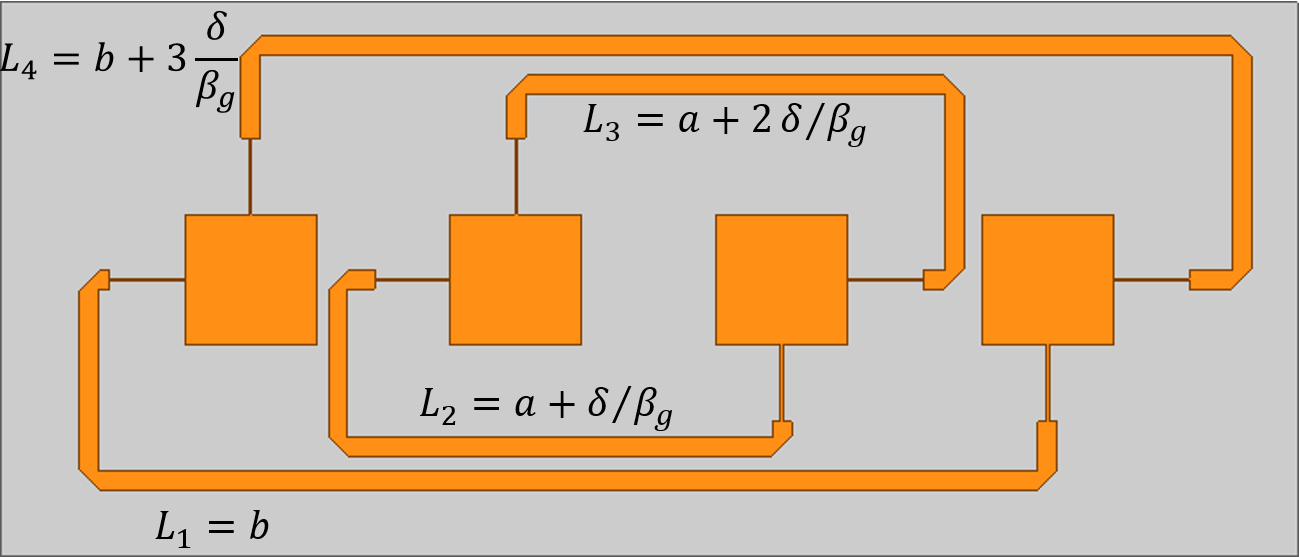}
	\caption{$\gamma=-15^{\circ}$ re-directing Van Atta array}
	\label{fig:VRPol4El} 
\end{figure}

To test the proposed principle, a polarized-duplexed, single 4-element row reflector with a $\gamma=\theta-\psi=-15^{\circ}$ offset angle was modeled at 10 GHz with ANSYS HFSS (Fig. \ref{fig:VRPol4El}). The square patch antennas measure 8.28 mm with a center-to-center spacing of 16.8 mm and matched to 1.21 mm wide microstrip lines with 0.15 mm wide quarter-wave transformers. These were modeled as 35 $\mu$m thick copper traces on a 35 x 82 mm, 0.5 mm thick Rogers RO3003 ($\epsilon_{r}=3$) substrate. Inter-connecting line lengths (including transformers) were initially sized to achieve retro-reflection: $a=48.08$ mm, $b=86.54$ mm. A planar design limited the range of line extension and $\gamma=-15^{\circ}$ was the widest offset angle achievable without overlap. The required phase gradient was calculated to be $\delta=52.5^{\circ}$, using (\ref{eqn:SteeringAngle}) with $\psi=5^{\circ}$ and $\theta=-10^{\circ}$. This was added by extending the connecting lines by multiples of $\delta/\beta_{g}=2.81$ mm, where $\beta_{g}$ is the guided propagation constant.

\section{Fabrication \& Experiment Setup}
\label{sec:prototype}
\begin{figure*}[!htb]
	\centering
	\begin{minipage}{.2\textwidth}
		\vspace*{\fill}
		\centering
		\includegraphics[width=\textwidth]{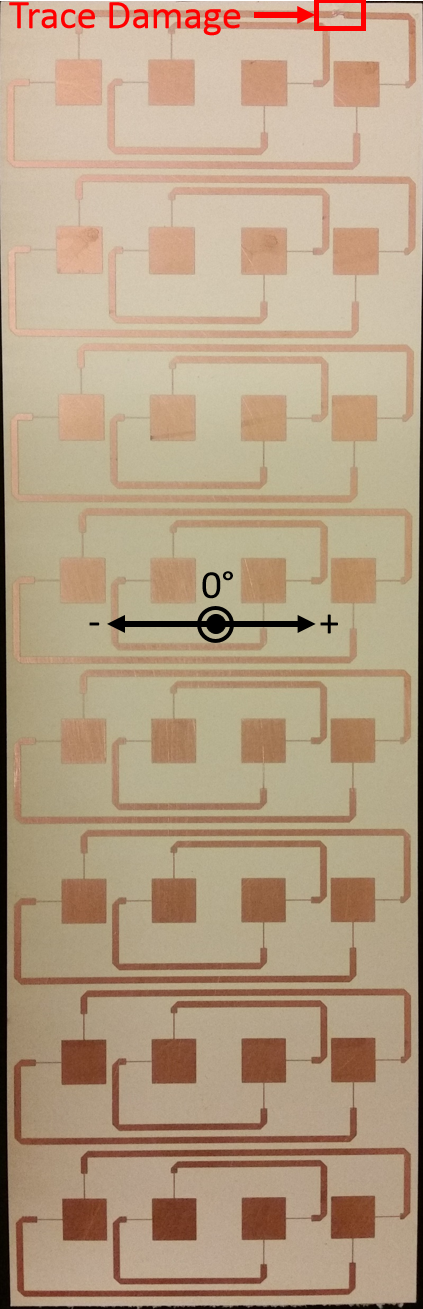}
		\subcaption{Fabricated re-director}
		\label{fig:MeasurementRfl}
	\end{minipage}\hspace{0.01\textwidth}
	\begin{minipage}{.66\textwidth}
		\vspace*{\fill}
		\centering
		\includegraphics[width=\textwidth]{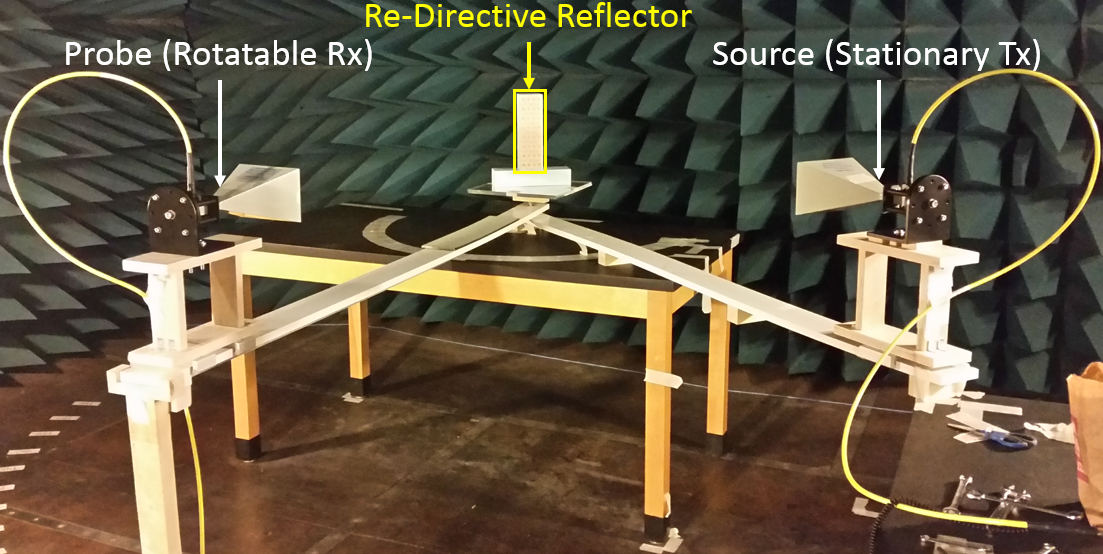}
		\subcaption{Experimental apparatus}
		\label{fig:MeasurementLayout}\par\vfill
		\includegraphics[width=\textwidth]{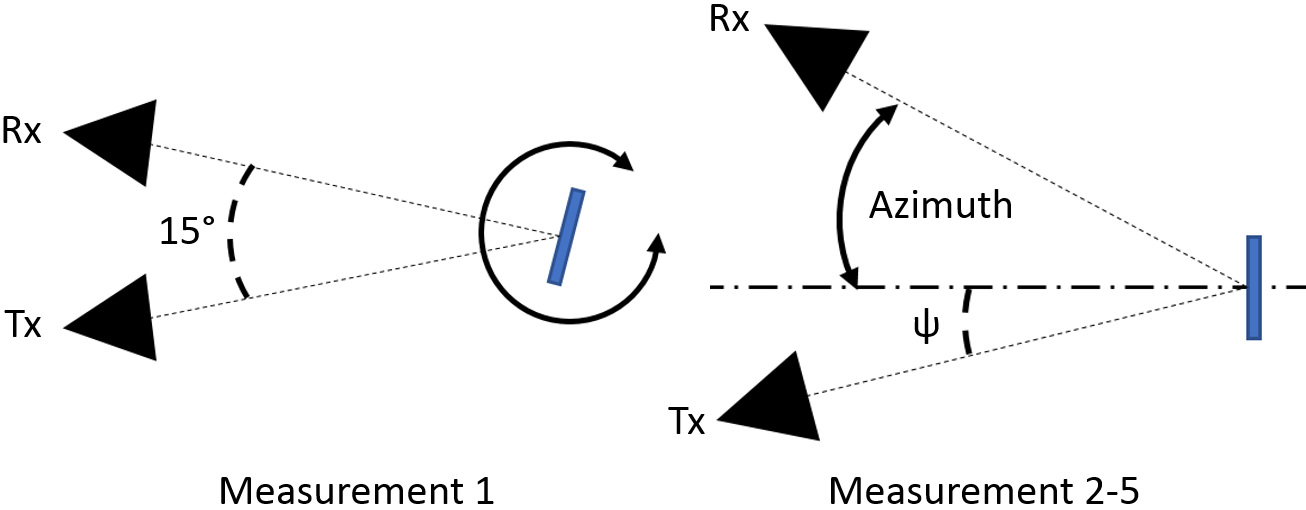}
		\subcaption{Measurement types}
		\label{fig:MeasurementTypes}
	\end{minipage}
	\caption{Fabrication \& experimental measurement}
	\label{fig:Measurement}
\end{figure*}

Real-world functionality was verified by fabricating and measuring an 8-row re-directing reflector. Each row was separated by a patch center-to-center spacing of 30 mm and printed on a 242.5 x 77 mm substrate. The fabricated reflector along with an angular reference is shown in Fig. \ref{fig:MeasurementRfl} where trace damage was noted. Measurements were performed using the bistatic apparatus in Fig. \ref{fig:MeasurementLayout}. A vertically polarized source horn was fixed to the right arm and kept stationary, while a probe horn was attached to a movable left arm. The orientation of the probe antenna could also be rotated to switch between measuring vertical (co-polarized) and horizontal (cross-polarized) polarizations. The following measurements were conducted over a frequency range of 8-12 GHz (Fig. \ref{fig:MeasurementTypes}):

\begin{enumerate}
	\item Fixed $\gamma=-15^{\circ}$ between source and probe, cross-polarized output measured across incidence angles
	\item Probe: bistatic sweep, Source: $\psi=30^{\circ}$ incidence
	\item Probe: bistatic sweep, Source: $\psi=15^{\circ}$ incidence
	\item Probe: bistatic sweep, Source: $\psi=-5^{\circ}$ incidence
	\item Probe: bistatic sweep, Source: $\psi=-15^{\circ}$ incidence
\end{enumerate}

Measurement 1 fixed the probe at a constant $-15^{\circ}$ offset from the source. The reflector was rotated by $5^{\circ}$ to position the source at a specific incidence angle where the probe would measure the re-radiated cross-polarized power at the designed $-15^{\circ}$ offset. This allowed reflector performance to be evaluated over a range of frequencies and incidence angles.

Measurements 2-5 fixed the source at a specific incident angle and then azimuthally swept the probe antenna at $5^{\circ}$ increments to obtain a scattered radiation pattern. Sweep resolution was reduced to $1^{\circ}$ within $\pm5^{\circ}$ of the expected specular and re-radiation angles. Three cuts were taken per measurement: Co and cross-polarized sweeps with the reflector to obtain specular and re-radiated patterns respectively and a co-polarized cut with an equally sized copper plate for a baseline. Sweep ranges and number of feasible incidence angles were limited by apparatus physical constraints. 

\section{Results \& Discussion}
\label{sec:results}
\begin{figure}[!b]
	\centering
	\includegraphics[width=3in]{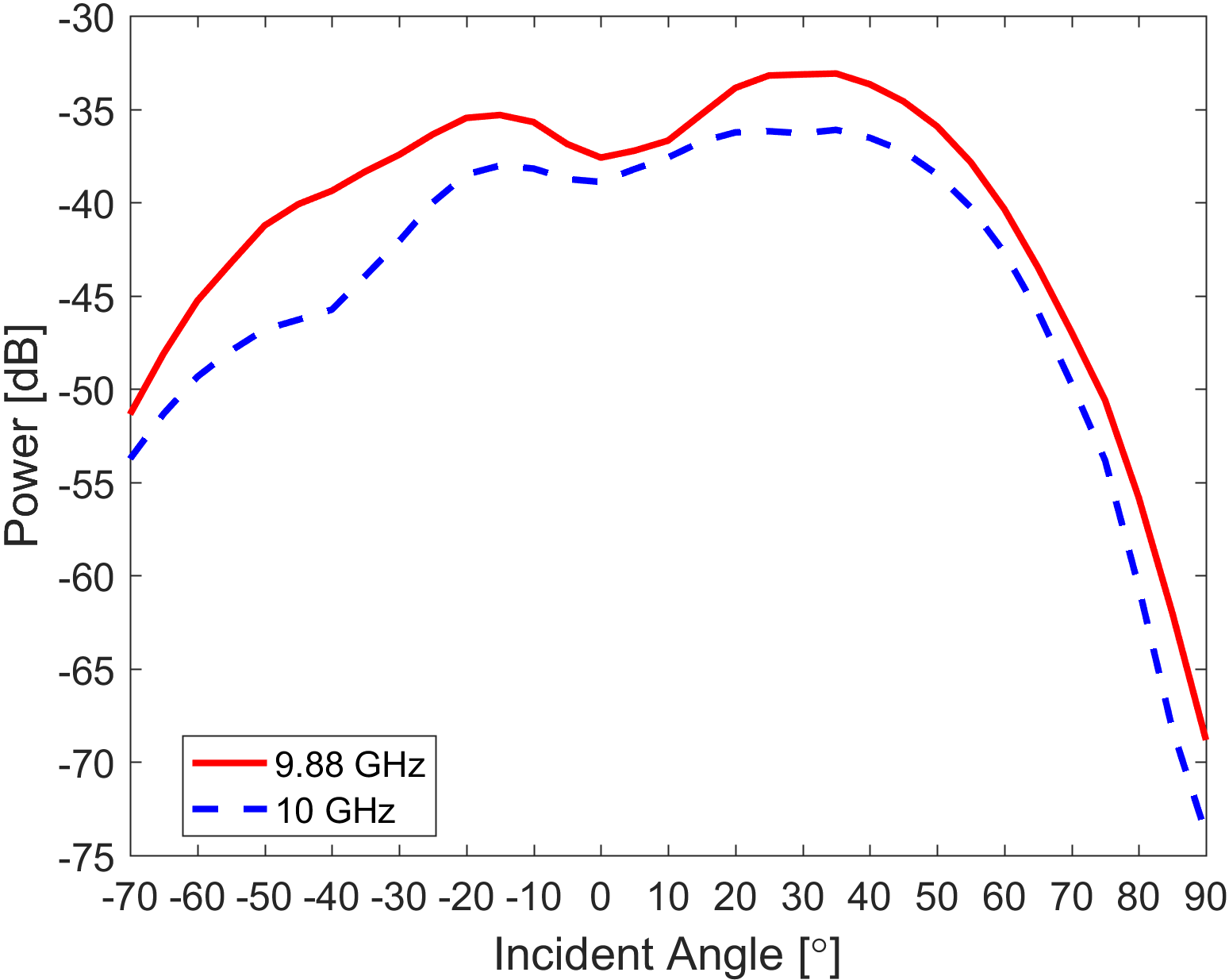}
	\caption{M1: Measured cross-polarized re-radiated power}
	\label{fig:FixedBistatic} 
\end{figure}

Fig. \ref{fig:FixedBistatic} plots the re-radiated power from Measurement 1 over a range of incident angles. The array's power output fluctuates slightly within $\pm(50-60)^{\circ}$ and drops off outside this envelope. These edge drop-offs are expected as conventional patch antennas cannot operate at extreme angles while the $\pm10^{\circ}$ fluctuation is attributed to resonance effects. Optimal performance is achieved at $30^{\circ}$ incidence as both the incident and re-radiated lobes are within the $\pm(50-60)^{\circ}$ envelope. In contrast, performance at negative incidence angles is slightly reduced since the re-radiated lobe is pushed to the $-50^{\circ}$ range. Furthermore, the fabricated reflector's optimal frequency shifted to 9.88 GHz; likely attributed to etching tolerances and naturally occurring variations in the substrate's material properties. Power outputs for both 9.88 and 10 GHz are plotted to highlight this difference in performance. 

\begin{figure*}[!htb]
	\centering
	\begin{subfigure}{0.4\linewidth}
		\includegraphics[width=\linewidth, height=0.9\linewidth]{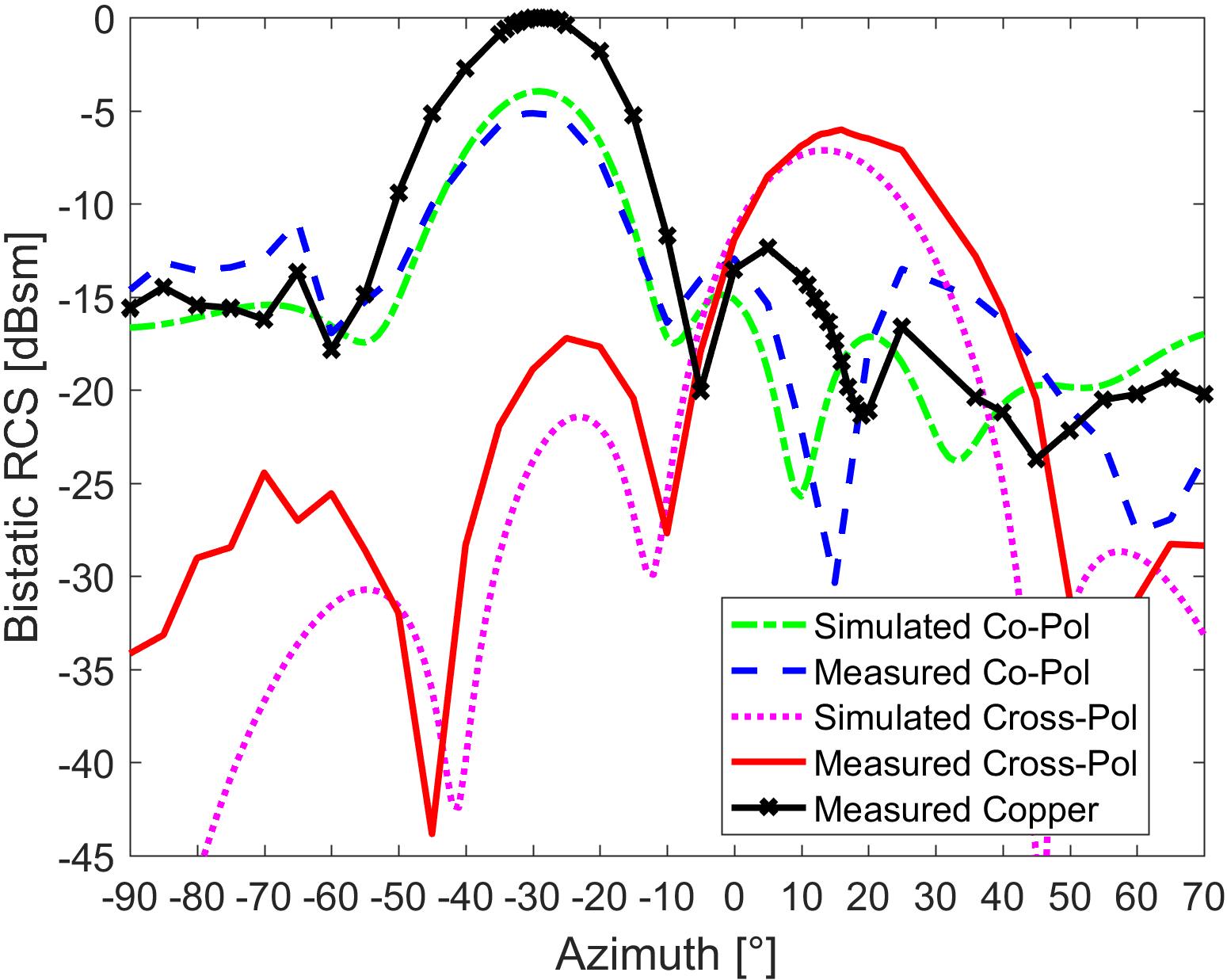}
		\caption{M2: $\psi=30^{\circ}$ incidence}
		\label{fig:Bistatic30}
	\end{subfigure}
	\begin{subfigure}{0.4\linewidth}
		\includegraphics[width=\linewidth, height=0.9\linewidth]{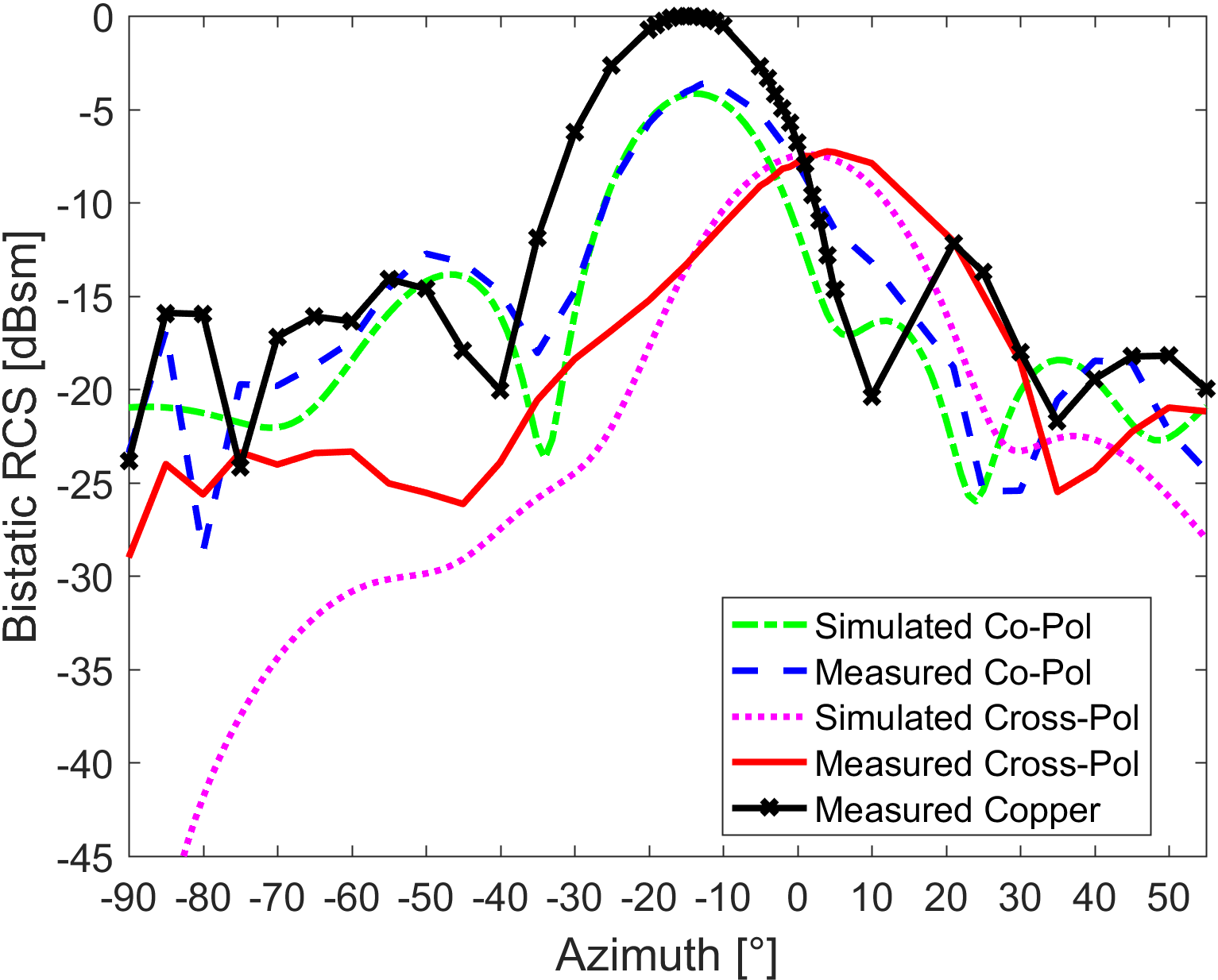}
		\caption{M3: $\psi=15^{\circ}$ incidence}
		\label{fig:Bistatic15}
	\end{subfigure}
	\begin{subfigure}{0.4\linewidth}
		\includegraphics[width=\linewidth, height=0.9\linewidth]{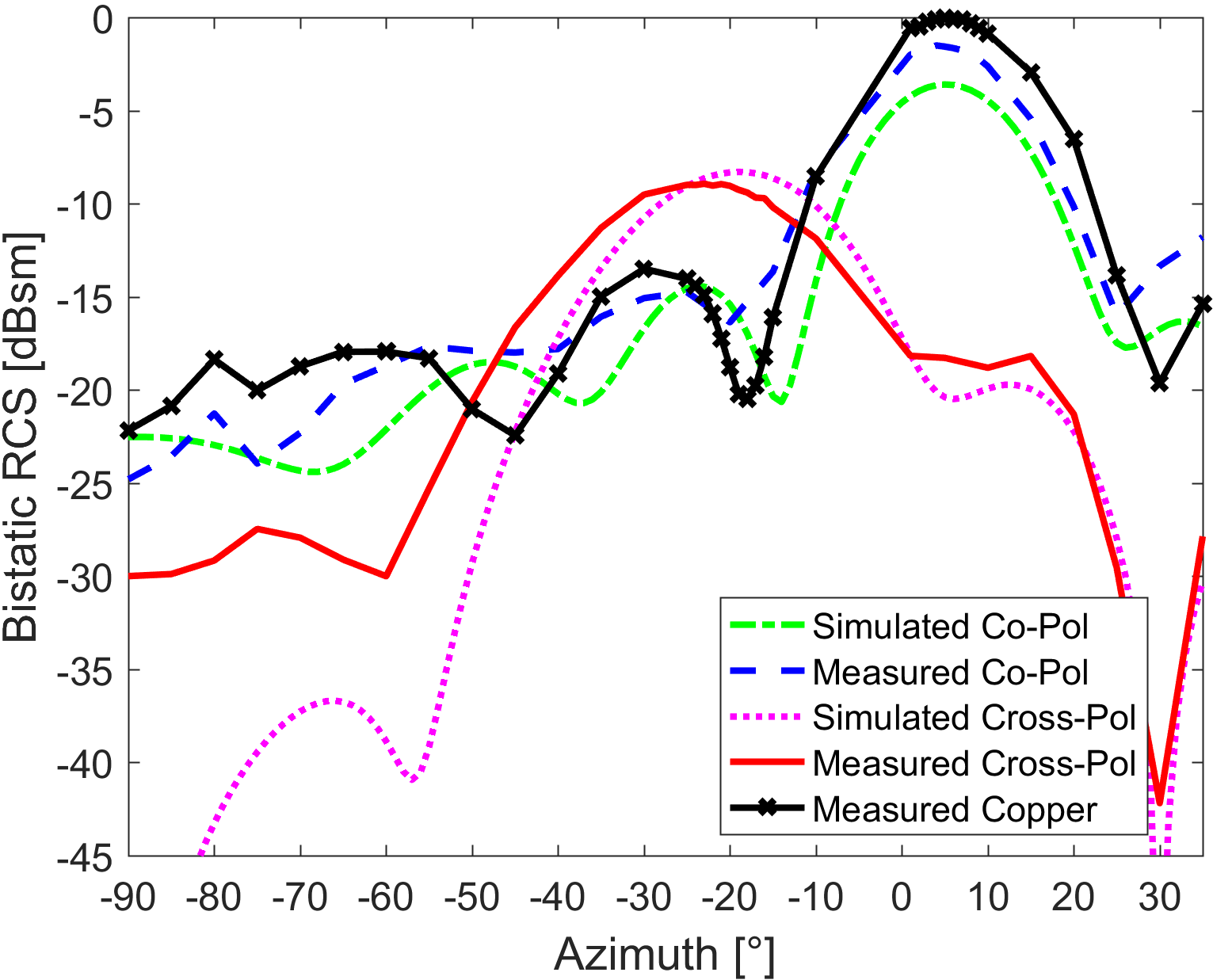}
		\caption{M4: $\psi=-5^{\circ}$ incidence}
		\label{fig:Bistatic_5}
	\end{subfigure}
	\begin{subfigure}{0.4\linewidth}
		\includegraphics[width=\linewidth, height=0.9\linewidth]{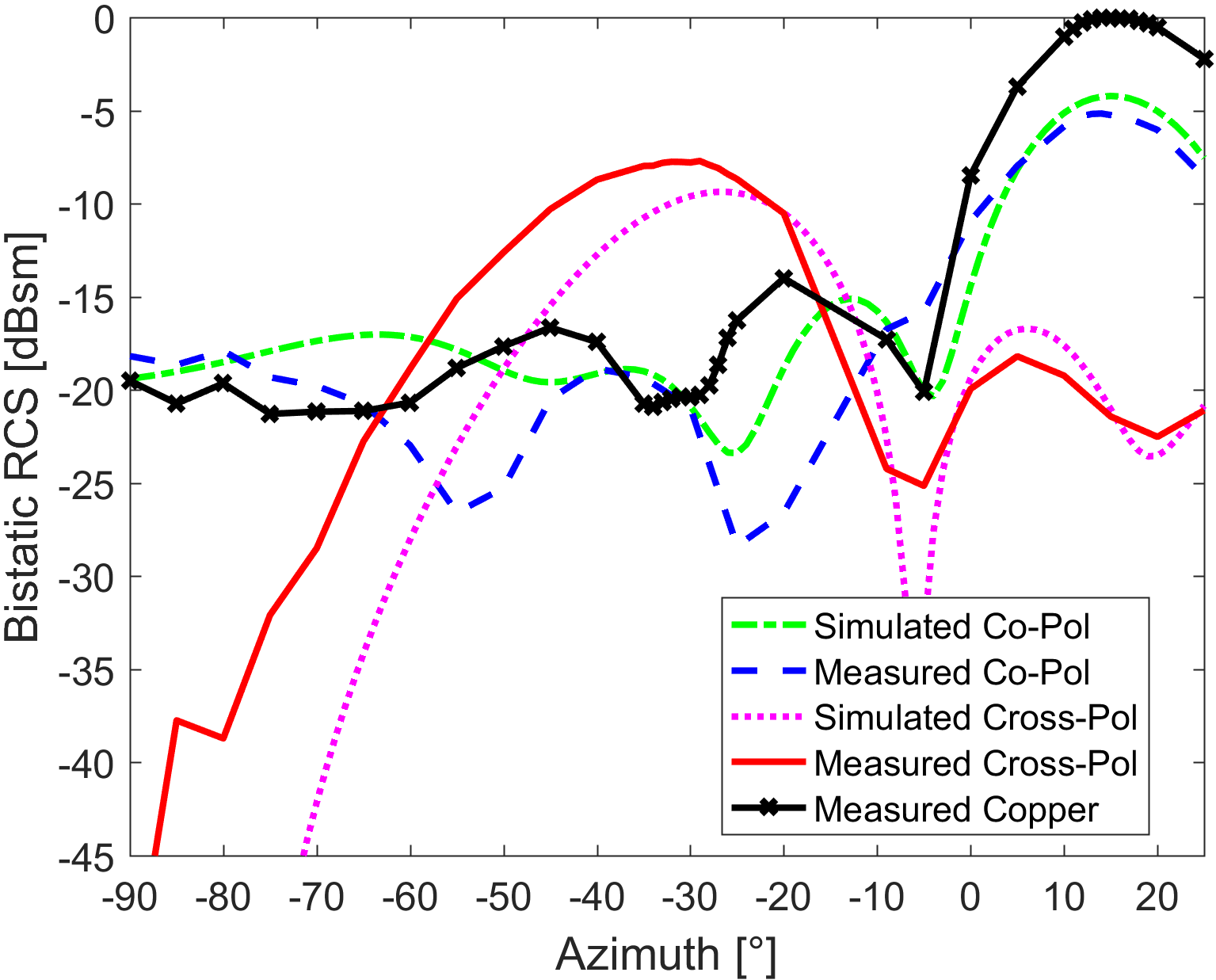}
		\caption{M5: $\psi=-15^{\circ}$ incidence}
		\label{fig:Bistatic_15}
	\end{subfigure}
	\caption{Normalized scattering patterns of $\gamma=-15^{\circ}$ re-directive reflector}
	\label{fig:BistaticRCS}
\end{figure*}

Fig. \ref{fig:BistaticRCS} plots the radiation patterns from Measurements 2-5 as bistatic RCS. Both simulated and measured RCS values are normalized to account for differences in reflector size, number of rows, and fabrication trace damage. These normalization patterns are obtained by simulating and measuring copper plates of the same dimensions as their re-directive counterparts. To reduce clutter, only the measured copper RCS is shown. For easier comparison, the radiation patterns are plotted at optimal frequency; 10 GHz for simulated and 9.88 GHz for the measured. Both the simulated and measured patterns show a re-radiated, cross-polarized main lobe approximately $-15^{\circ}$ from incidence, demonstrating functionality over multiple incidence angles. Re-direction accuracy falls within $\pm5^{\circ}$ and can be attributed to the non-linear nature of (\ref{eqn:SteeringAngle}). In particular, the required gradient is calculated from a specific incident and re-radiation angle pair rather than a specified offset. Consequently, minor drifting may occur at other incident angles. Due to antenna and transmission line properties, some power is specularly reflected by the array. The polarization-duplexing makes it easy to separate these specular components as the re-radiated main lobe is cross-polarized. Absorption of incident power is evident as the (co-polarized) specular reflection lobes of the arrays are lower than those of the copper plate. Furthermore, measured array co-polarized power is inversely proportional to cross-polarized power, suggesting that any power not received is specularly reflected. As shown in Fig. \ref{fig:FixedBistatic}, optimal performance occurs at $30^{\circ}$ incidence where measured re-radiated power is approximately equal to the specularly reflected power. The worst performance occurs at $-5^{\circ}$, where the measured peak specular reflection is 7.4 dBsm higher than the re-radiated peak. At all other incidence angles, the specular/re-radiation difference is approximately 3-5 dBsm. 

Array performance can be improved by increasing re-radiated power or reducing specular reflection. The former is dependent on element quantity and matching while minimizing exposed non-radiating surfaces reduces the latter. Trimming the upper left and lower right corners of the model in Fig. \ref{fig:VRPol4El} reduced the simulated specular/re-radiation difference by up to 3 dB over all incidence angles. 

\section{Conclusions}
\label{sec:conclusion}
In this Letter, array theory was used to passively modify the re-radiation angle of a Van Atta retro-reflector. Unlike other passive re-direction techniques, this method effectively applies the same angular offset over a wide range of incident angles at a fixed frequency. Further verification was provided through simulation and experimental measurements on a $-15^{\circ}$ offset re-directive reflector. Polarization-duplexing was used to passively and reciprocally achieve the required isolation in the interconnecting networks. While the principles behind the modified Van Atta array are universal, reflector performance was limited by the properties of the planar antennas and microstrip lines used. In particular, the amount of power received and re-radiated decreased significantly at steeper incident angles. Moreover, peak specular levels were 0.8-7.4 dBsm higher than their re-radiated counterparts over the tested incidence angles.

Future work is expected to focus on improving reflector performance and angular range. These can be accomplished by strategies such as using wider accepting-angle antennas, increasing elements, and decreasing exposed non-radiating surfaces. In addition, variable phase shifters can be added to the interconnecting lines, in place of line length adjustments, to enable variable beam steering and tuning.


\IEEEtriggeratref{3}
\printbibliography[title={References}]

\end{document}